# Recent Conceptual Consequences of Loop Quantum Gravity
# Part III: A Postscript On Time

Rainer E. Zimmermann

IAG Philosophische Grundlagenprobleme,
FB 1, UGH, Nora-Platiel-Str.1, D – 34127 Kassel /
Clare Hall, UK – Cambridge CB3 9AL[1] /
Lehrgebiet Philosophie, FB 13 AW, FH,
Lothstr.34, D – 80335 München[2]
e-mail: pd00108@mail.lrz-muenchen.de

### Abstract

With a view to the results discussed in the first two parts of this paper, the concept of time is revisited oncemore and chosen as an example in order to demonstrate the meaning of „fundamental" in both physical and philosophical terms. It is shown that if arguments are given in favour of or against the concept of time, then their ontological state has to be clarified in the first place.

### Introduction

In the two preceding parts of this paper, aspects of modern metaphysics with a view to consequences of recent results in loop quantum gravity have been discussed in some detail, under both a foundational [1] as well as holistic [2] perspective. It has been found that on a truly fundamental level, the world turns out to be beyond traditional framework categories such as space and time. The important point is however, to actually identify that fundamental level, in the first place. The terminology of philosophy and physics is different here so that it is necessary to give a sound definition of the „foundations" of the world. According to a tradition which refers back to philosophers such as Spinoza, Leibniz, and others, we have noticed that (re-phrasing Schelling in fact) foundation is against that to which it is foundation, *non-being*. Hence, it is to the physical world we (empirically) observe what possibility is to actuality. Space and time therefore (as well as matter), are concepts which have no ontological meaning beyond the physical world. With respect to the truly fundamental level (non-being), they are emergent properties of the world to which this level is fundamental. As to time in particular, the foundation of the world is „eternal" in the strict sense, not referring to an infinite time length, but to a complete *absence* of time instead. Nevertheless, the conception of visualizing time (if not also space) as something which is available on the fundamental level of the world is still under dispute. So recently, Smolin has discussed what he calls challenges to the arguments aiming at the elimination of time. [3]

---

[1] Permanent addresses.
[2] Present address.

As a postscript to the first two parts of this paper, we will discuss here a) the question as to from what position the arguments in favour of the elimination of time actually do argue, and b) what it is therefore, that Smolin defines as challenges of that argumentation. We start first of all with recalling some of Barbour's arguments as he has explained them in his book. [4]

# 1  Barbour's Conception Revisited

Barbour visualizes the *physical* (quantum) world as a static world onto which time is being projected by human observers. In a sense, there is a collection of possible states of the world, and the human brain composes a movie of a sequence of states selected from this collection. Hence, the points of a world of such Nows are worlds unto themselves: „No thread of time joins them up." ([4], p.45) A given situation as it is being signified by some observer as having developed historically, can be thought of as a *time capsule*, i.e. a single configuration of the Universe „that seems to be the outcome of a dynamical process of evolution through time in accordance with definite laws." ([5], p.408) Hence, the meaning of *law* here refers to rules which tell us how to transform configurations from being possible to becoming actual (to establish, as Barbour says, which configurations go from the „heap" of possibilities into that of actualities). ([5], p.409) Alternatively, the respective probability distribution for one or the other transition of this kind can be visualized as distribution of a kind of „mist" such that the appearance of time arises from the fact that the mist is concentrated on time capsules. Hence, the likeliness to experience a Now that is a time capsule is greater than to experience one which is not. ([4], p.52) In fact, this can be equivalently discussed in terms of the celebrated Wheeler-DeWitt equation of the form $H\Psi = 0$, where H is an appropriate Hamiltonian operator, and $\Psi$ is the „wave function of the Universe".

I have criticized this approach earlier in the following sense: The idea that time emerges out of timelessness, is not new as far as philosophy is being concerned. Hence, Barbour's sample world, called „Platonia", should be named „Spinozania" instead. (This is so because substance encodes structures in its potentiality, but not concepts humans develop with respect to these structures.) Time shows up as a convention, i.e. as part of human communicative strategies, rather than as something which is a generic property of the Universe. It is only property of the human world (of the physical Universe that is), in so far as there is a convention which turns out to be practical for human life, if conceptualized according to what humans perceive of their world. The actual choice is twofold then: Either we agree that humans perceive their world in some incomplete manner.

Then they model their world according to this incomplete perception. Or we agree that their mode of perception defines their worldliness. Then they model their world completely. But there may be many worlds. In the one case, humans perceive the one (unique) *real* world under their generic (modal) *perspective*. Hence, there may be many such perspectives. In the other case, humans perceive *their own (modal) world* as it is. But the *real* world is a collection of the many worlds. Recalling Spinoza's conception of the one substance and its infinitely many attributes, we realize that these two cases are essentially equivalent as far as epistemology is concerned, but different from the ontological point of view. There is no crucial experiment one could give in favour of the one or the other. But the speculative consequences are manifold.

The point is that we have to carefully separate the physics from the philosophy. Barbour formulates: „If we could see the universe *as it is*, we should see that *it is static*. Nothing moves, nothing changes." ([4], p.39, emphasis mine) In strict terms, this is not quite correct: The expression „static" implicitly refers to a convention of physics, namely to call something static for which the time variable is absent in the relevant functions. But substance is not static, because it is not a reduction of some function to another one for which the „parameter" t can be set to zero or does not show up at all. This is so because „t" *was never present*, in the first place. Hence, the term „static" refers to physics, not to philosophy. On the other hand, the formulation of „the universe *as it is*" refers to the case of the *real* Universe. But this is not empirically attainable for humans who can visualize the Universe in *modal* terms only. Hence, „it is" is a formulation which refers to philosophy, not to physics. Consequently, to be precise, the formulation should rather be of the form: „If we could perceive (or measure) the universe as it is *realiter* [but what we cannot actually do], we should notice that there *is* no time." [Only *modaliter* therefore, do we have the impression of a world which appears *as if there were* time. And we speak and behave accordingly.]

Note by the way though that even *within* the world, time perception is far from being settled. Instead it is highly contextual. The more so, as we know by now that even in everyday terms, we cannot be sure about our temporal perception. As Eagleman and Sejnowski have shown recently [6], human temporal perception refers to a generic „time window" of about 80 msec which is utilized by the brain in order to co-ordinate sensory input data belonging to one and the same event, but arriving in a sequential manner at the cerebral stimulus centre due to their different channels (and speeds) of propagation. Basically, for being able to synchronize all of them thus producing one perceived event, the brain has to actually backdate the result. In other words: Even with respect to common everyday perceptions (pain e.g.) do we live „in our past". This is already very much on the line of what Barbour calls the „movie" our brain composes of all the incoming sensory perceptions, and to a lapse function which is being created by this procedure of composition. The consequences of this for the human means of constructing their (everyday) world has been discussed in more detail in [7].

Hence, coming back to the fundamental concept of time, the dynamics described by Barbour is essentially one of 3-dimensional Riemannian spaces sequentialized in terms of an ordering principle. This idea going back to John Wheeler in the sixties is discussed in Julian Barbour's book, where he points out that the „key geometric property of space-times that satisfy Einstein's equations reflects an underlying principle of best matching built into the foundations of the theory." ([4], p.176) The time separation of spatial slices shows up here as what Barbour calls a *distinguished simplifier*, as an ordering principle for making unfoldings simple. ([4], p.180) If time is being visualized as a mere ordering principle, then, in philosophical terms, we are left with space as an attribute. Note however, that the *dimensionality* of space is only a finite representation then, which does not reflect the true nature of space, but only our modal attitude towards it with a view to spatial ordering. This has been shown earlier [8] with respect to the approach of topological quantum field theory which points to a direct correspondence between the change of spin numbers in spin networks (on the „microscopic level" of description) and the change of space topology (on the „macroscopic level" of description). Note again that „time" showed up here not as a function, but as a manifold. And this is particularly interesting, because with a view to what Barbour tells us about the „absence" of time, this means that the concept of time is *intrinsically included* here as a pragmatic ordering principle for localizing topology changes. This is similar to what Prigogine calls the „age of a system", which is roughly a frequency of formations of new structures in a system making the latter more complex. Time *as a convention* then, would be an approximate „average" over such ages. Hence, time shows up as being associated to a kind of measuring device for local complexity gradients. So what we have in the end, is a rough (and simplified) outline of the foundations of emergence, in the sense that we can localize the fine structure of emergence (the rearrangements of spin numbers in purely combinatorial terms being visualized in philosophical terms as a *motion-in-itself*) and its results on the „macroscopic" scale (as a change of topology being visualized by physical observers as a *motion-for-itself*). This is actually what we would expect of a proper theory of emergence. But note also that space and time, in the classical sense, are obviously absent on a fundamental level of the theory, although they can be recovered as concepts when tracing the way „upward" to macroscopic structures. In other words: even as a gross average feature for „shortsighted" human scientists (as Penrose indicates it at the end of his first twistor paper), space and time would nevertheless turn up as (philosophical) categories of concepts, simply, because the meaning of these categories is well-adapted to what humans actually perceive of their world (and communicate to other humans). This is in fact, a point, where Barbour's argument seems to break down (if discussed within this philosophical perspective): What he essentially shows in his book is that quantum theory, *in so far as it is foundational*, describes partly what was called non-being (or substance) in former times. Hence, there is neither space nor time in *real* terms ( = realiter, i.e. with respect to what there is in an absolute sense of the world's foundation), but there *is* space and time in *modal* terms ( = modaliter, i.e. with respect to what humans perceive of their world). The former refers to substance, or, alternatively, to what *speculative* philosophy is all about. The latter refers to the physical world, or, to what *sceptical* philosophy is all about. The one relies on theoretical speculation according to what we know - speculating about the foundation of the world, which is outside (logically „before") the world, and of which we are not a part therefore, and hence, about which we cannot actually know anything. The other refers to the empirical world, about which, with the help of experiments, we can obtain knowledge, in fact. Obviously, in terms of physics, the first (speculative) aspect is corresponding to physical theory, in so far as it is foundational. The second (sceptical) aspect corresponds to physical theory, in so far as it is empirical.

## 2 Smolin's Arguments

So what we realize is that communicational difficulties between physics and philosophy arise when the demarcation of their respective fields of activity remains unclear. To make this more precise has been one task of this present paper. The difficulty of interdisciplinary co-operation is that each participant is asked to give up his/her own prejudices, in the first place, because „interdisciplinary" does not refer to either discipline, but to something which is „between" the disciplines, and which is something entirely new, therefore. (Similar to the connotation of „intercultural" which does not refer to any of the various cultures, but to this innovative „in between".)

Hence, checked against this attitude, we can easily realize that the arguments which are listed by Smolin [3] in order to challenge Barbour's conception centre around physical epistemology, but do not conceptualize the notion of time with a view to precise differentiation of „world" and „foundation". So his starting point is already one which neglects the aforementioned demarcation. He states „that every observable in a theory of cosmology should be measurable by some observer inside the universe" and that „all mathematical constructions necessary should be realizable in a finite time." This is certainly a useful assumption, if the reasonable modelling of physical phenomena is concerned, but it does not carry any additional philosophical connotation so that the further statement: „However my view is that this question [of the absence of time] is not one that can be settled by philosophical argument alone." is not mediated with the epistemological starting point. (Besides the fact that there is probably no reasonable philosopher who would seriously undertake „to settle this question" by himself.)

In fact, the question is not (and we have discussed this by now in some detail) whether measurements *within* the physical world are limited somehow in terms of information to be gained (or not). Or whether, in principle, time could be eliminated from theories *about that world*. The fact is: there *is time* (at least in epistemological terms of a utilized concept) down to the basic quantum level of theories. The real question is instead: Is there time *at all* in the *real* world (of which we can only observe a tiny part in principle)?

Smolin also emphasizes the role played by the requirement „that a *theory of cosmology* must be falsifiable in the usual way ... This leads to the requirement that a sufficient number of observables can be determined by information that reaches a *real observer inside the universe* to determine either the classical history or quantum state of the universe." (emphasis mine) Hence: „If we want to do cosmology, we must restrict ourselves to theories in which all observables are accessible to real observers inside the universe." Again, the oberserver who is a part of the Universe, cannot be a *real* observer, but rather a *modal* one. In fact, there are no observers except modal ones. (They are not *inside* the Universe though, because their perception actually defines the structures which constitute the Universe. Instead, they are an intrinsic part of the Universe.)

Similar conceptual problems we can find in other passages: Smolin quotes e.g. Markopoulou who argues in favour of an „interconnected web of Hilbert spaces tied to the causal structure such that each act of observation [considered as an event] is represented in terms of a Hilbert space constructed to represent the information available." Obviously, this is a formal problem of physical epistemology, but does not tell anything about the foundational aspects. (Because there *modally is* a causal structure so long as we have space and time (quanta), i.e. down to the spin network level. But what about the foundation of spin networks?)

The same problem occurs with the information transport within the world: „It is only by insisting that the context of real observers inside their universe is defined by the information that reaches them by means of radiation that propagated from their past that a link is made between the issue of observability of the universe and its causal structure." True for physics. But obviously, if on the truly fundamental level, there is no time (and space), the propagation of radiation is ill-defined. Hence, the „projective mechanism" (of what we might call the functor Past) has its starting point in the classical world (on the level of humans that is). Back projection comes first, propagation of information from the past comes second. (Note that this *does not mean* that humans would arbitrarily construct phenomena which they define afterwards. Contrary to that sort of solipsism, we assume that *there is something* real (independent of human perception), but how it really „looks" cannot be perceived.)

The point is that human observers conceptualize the world as if it would be constituted of phenomena whose information is propagated through space and time. So if for Smolin, it appears as being „hard to divorce the notion of causal structure from a finite speed for the propagation of information and hence time." he is quite correct. But this does not tell us anything about the real world, it simply characterizes the perception of the modal world.

So what we realize is that it is essentially the unclear demarcation of epistemological and ontological problems (i.e. of physical and philosophical questions) that creates misunderstandings in the discussion of basic framework categories such as space, time (and matter). This is demonstrated again when Smolin comes to his conclusions: „If the universe is discrete and time is real, and is itself composed of discrete steps, then time may be none other than *the process which constructs*, not only the universe, but the space of possible universes relevant for observations made by local observers. / Beyond this, there is the possibility of a quantum cosmology in which the actual history of the universe up till some moment and the space of possible universes present at this ‚instant' are not two different things, but are just different ways of seeing the same structure whose construction is the real story of the world."

It is difficult to see how time (whether it is visualized in real or modal terms) can be a process. (This appears to be a problem of lexicology rather than of physics or philosophy.) And as it is a local concept utilized by observers, how can it construct the Universe altogether? And what about „the space of possible universes"? The foundation of the world (possibility as opposed to actuality) has not been defined in the paper at all (as far as I can see). So what concept of possibility is here utilized all of a sudden? Finally: How can „the actual history of the universe" (as re-constructed by the modelling of some observer) and „the space of possible universes" be one and the same thing of which different aspects can be observed (by whom)?

## 3   Conclusion

Hence, our main result here is the following: Barbour, in so far as he addresses the physical problem of time, discusses the problem of whether quantum theory could be formulated without a time variable. But as far as he discusses (as is his declared intention) the question whether there is time in principle (in real terms, not in modal terms), he addresses a philosophical problem. As to the first, he raises interesting questions. As to the second, he does not tell us anything new. (Hence, the title of his book is somewhat exaggerating.) Smolin however, in so far he addresses the physical problem of time, can list a number of challenges to the problem of whether quantum theory could be formulated without a time variable. But as far as he discusses the question whether there is time in principle, addressing therefore a philosophical question, he does not give any philosophical argument. Hence, as to the first, he also raises interesting questions. But as to the second, he does not only tell us nothing new, but nothing at all, so that his conclusions are more than unclear.

Apparently, the problem is that many physicists have realized by now that their theories need philosophical conceptualization. This is actually an old idea dating back to a famous talk given by René Thom some time ago. [9] And in the meantime, there is a number of philosophers around who have also some knowledge of physics. And in fact, there have been occasions by now to enhance the communication among the disciplines. One example can be found in [10]. But innovative results from such interactions can only be expected, if the various oppositions which have emerged during the long history of separation which has characterized the relationship between physics and philosophy (as well as the relationship of schools *within* physics and philosophy, respectively) for the last two hundred years can be successfully dissolved. Hence, communication should replace confrontation.